# Evidence for thermal activation in the glassy dynamics of insulating granular aluminum conductance

**T Grenet and J Delahaye**
CNRS, Institut NEEL, F-38000 Grenoble, France
Université Grenoble Alpes, Institut NEEL, F-38000 Grenoble, France

E-mail : thierry.grenet@neel.cnrs.fr

**Abstract**: Insulating granular aluminum is one of the proto-typical disordered insulators whose low temperature electrical conductance exhibits ubiquitous non-equilibrium phenomena. These include slow responses to temperature or gate voltage changes, characteristic field effect anomalies and ageing phenomena typical of a glass. In this system the influence of temperature on the glassy dynamics has remained elusive, leading to the belief that the slow relaxations essentially proceed via elastic quantum tunneling. A similar situation was met in insulating indium oxide and it was concluded that in high carrier density Anderson insulators, electrons form a quantum glass phase. In this work we experimentally demonstrate that thermal effects do play a role and that the slow dynamics in granular aluminum is subject to thermal activation. We show how its signatures can be revealed and activation energy distributions can be extracted, providing a promising grasp on the nature of the microscopic mechanism at work in glassy Anderson insulators. We explain why some of the experimental protocols previously used in the literature fail to reveal thermal activation in these systems. Our results and analyses call for a reassessment of the emblematic case of indium oxide, and question the existence of a quantum glass in any of the systems studied so far.

## 1. Introduction

Glassy dynamical features have been studied in a variety of disordered insulators and interpreted as signatures of the electron glass. Recent reviews of the experimental and theoretical works may be found in [1]. It has been shown that these systems are never in equilibrium at low temperature and proceed infinitely slowly to an equilibrium state. The relaxations are recorded by measuring the low temperature electrical conductivity, a quantity which is generally very sensitive to these insulators internal characteristics (e.g. amount of disorder or carrier density). The ultimate equilibrium state is believed to possess the minimal conductance and one indeed observes that after a quench from high to low temperature the conductance always logarithmically decreases with time, without any observable saturation on accessible experimental time scales. Field effect measurements performed on gated samples show characteristic conductance minima ("memory dips") located at gate voltages under which the samples have been allowed to relax for some time, an example of which is shown in the inset of Figure 1. The memory dip at a given gate voltage is slowly erased in a characteristic way after the gate voltage has been shifted to another value (see below).

In most glasses, like structural or spin glasses, the dynamics is thermally activated and slows down drastically as the temperature is decreased to the glass transition temperature. However in disordered insulators no signature of a glass transition at finite temperature was found, and in indium oxide and granular aluminum, two extensively studied systems, no signature of thermal activation of the dynamics was observed, leading to the belief that these might be quantum glasses where the slow relaxation proceeds mostly via quantum tunneling [2]. In recent years the glassiness and its associated anomalous field effect have been studied in other systems. The unusual temperature dependence of the conductance relaxations found in thin films of discontinuous metals [3] suggests a slow down of the dynamics upon cooling, and in amorphous $Nb_xSi_{1-x}$ alloys [4] the activated character of the dynamics was directly demonstrated. In light of these recent findings we re-examine the question of the elusive temperature dependence of the dynamics in insulating granular Al films. Our new experiments show unambiguously that the slow dynamics does strongly depend on temperature and is progressively frozen when the samples are cooled, thus dismissing the existence of a purely quantum glassiness in this system. These results and the analysis of previously used experimental protocols in the literature also put the existence of the quantum glass phase in other disordered insulators into question.

The present paper is organized as follows. The next section is devoted to the experimental protocol used. In section 3 we present the experimental results demonstrating thermal activation and clarify why previous experiments led to erroneous conclusions. In section 4 we discuss the activation energies involved in our system, comment on the case of indium oxide, and give some perspectives opened by our findings.

## 2. Experimental protocol

When studying the slow dynamics of a disordered insulator one always faces the same difficulty: the logarithmic conductance relaxation, which corresponds to the growth of a new memory dip, shows no characteristic relaxation time which temperature dependence could be directly studied. Moreover one cannot simply compare the slopes of the relaxation curves for different samples or for a single sample at different temperatures. This is because the mechanism through which the slow relaxations alter the conductance is not understood quantitatively. Hence comparing different curves, one does not know if different slopes reflect different dynamics or different sensitivities of the conductance to the changes occurring in the samples. For instance more conducting samples have significantly lower $\frac{1}{G}\frac{dG}{dLn(t)}$ slopes than more insulating ones, but there is no reason to believe that the relaxations are much slower in conducting samples than in highly insulating ones.

Actually we showed earlier [5] that due to ageing effects the slow growth of a new memory dip is not always exactly logarithmic in time. The observed departures from Ln($t$) may be used to study the $T$ dependence of the dynamics, however these are small and it may not be very convenient to do so.

In attempts to circumvent the difficulty, protocols using the comparison of two memory dips have been used in the literature. But as we showed earlier [6] they can be problematic and lead to wrong conclusions. For completeness we briefly review and comment on them in Appendix A.

The present study involves an "erasure protocol". It is free of the problems present in other protocols and gives reliable results. The basic idea is quite natural: in order to search for a temperature dependence of the dynamics, one has to "activate" a given set of relaxation times of the system, change the temperature, and re-measure them to see whether they have changed or not. In practice we build a memory dip at a given temperature $T_{build}$ and measure the time needed to erase it after the temperature has been changed to $T_{erase}$.

The procedure is as follows. A device designed as a MOS-FET, where the channel is the disordered insulator under study, is cooled from high temperature (typically larger than 100K) to $T_{build}$ under a cooling gate voltage $V_{g\ cool}$. For the building step the gate voltage is shifted to $V_{g\ build}$ and is kept at this value for the building time $t_{build}$ so that a memory dip slowly develops. Roughly speaking this step involves a set of relaxation times distributed between a lower cutoff value and $t_{build}$. Then the gate voltage is shifted to another value $V_{g\ erase}$, and one monitors the rate at which the memory dip at $V_{g\ build}$ is erased, that is to say the rate at which the system looses the memory of its relaxation under $V_{g\ build}$. Examples of $V_g$ scans performed to monitor the erasure are shown in Figure 2, where one sees the progressive erasure of the memory dip at $V_{g\ build}$ =0V while $V_g$ is kept at $V_{g\ erase}$=+15V. If the protocol is *isothermal* (no temperature change, $T_{build}$ =$T_{erase}$) one obviously expects that it takes roughly $t_{build}$ to erase the memory. If the protocol in *not isothermal* ($T_{build} \neq T_{erase}$), the distribution of relaxation times may change, a thermal activation giving rise to a slower (resp. faster) erasure if it is performed at a lower (resp.) higher temperature than $T_{build}$.

A simple analysis shows that in the isothermal erasure protocol, one expects the amplitude of the memory dip to be erased like [7, 8] :

$$\delta G(t) \propto Ln\left(1+\frac{t_{build}}{t}\right) \qquad (1)$$

This analysis relies on a relaxation time distribution $P(\tau_i)\sim 1/\tau_i$, corresponding to purely logarithmic relaxations.

Actually things are a bit more complicated. We have previously shown [5] that Eq. (1) is only observed if one "ages" the sample for a sufficiently long time at $T_{build}$ between the cool down and the building step. Indeed the ageing effects mentioned above (departures from exactly logarithmic relaxations) modify the erasure dynamics. Generally speaking the system's response, hence its effective relaxation time distribution $P(\tau_i)$, depends on its total age since the cool down, a hallmark of a true glass phase. The erasure function then $\delta G(t)$ differs from $\mathrm{Ln}(1+t_{build}/t)$, and the characteristic erasure time that one can define from it exceeds $t_{build}$ and can be as large as $3t_{build}$. This phenomenon exists in the present study, however since we will be interested in much larger deviations caused by temperature changes, we will neglect it (this effect has its counter part in spin glasses, and as far as we know still awaits a quantitative explanation).

To summarize isothermal erasure protocols give roughly Eq (1) and in non isothermal experiments, the existence of thermal activation should produce large deviations from Eq (1).

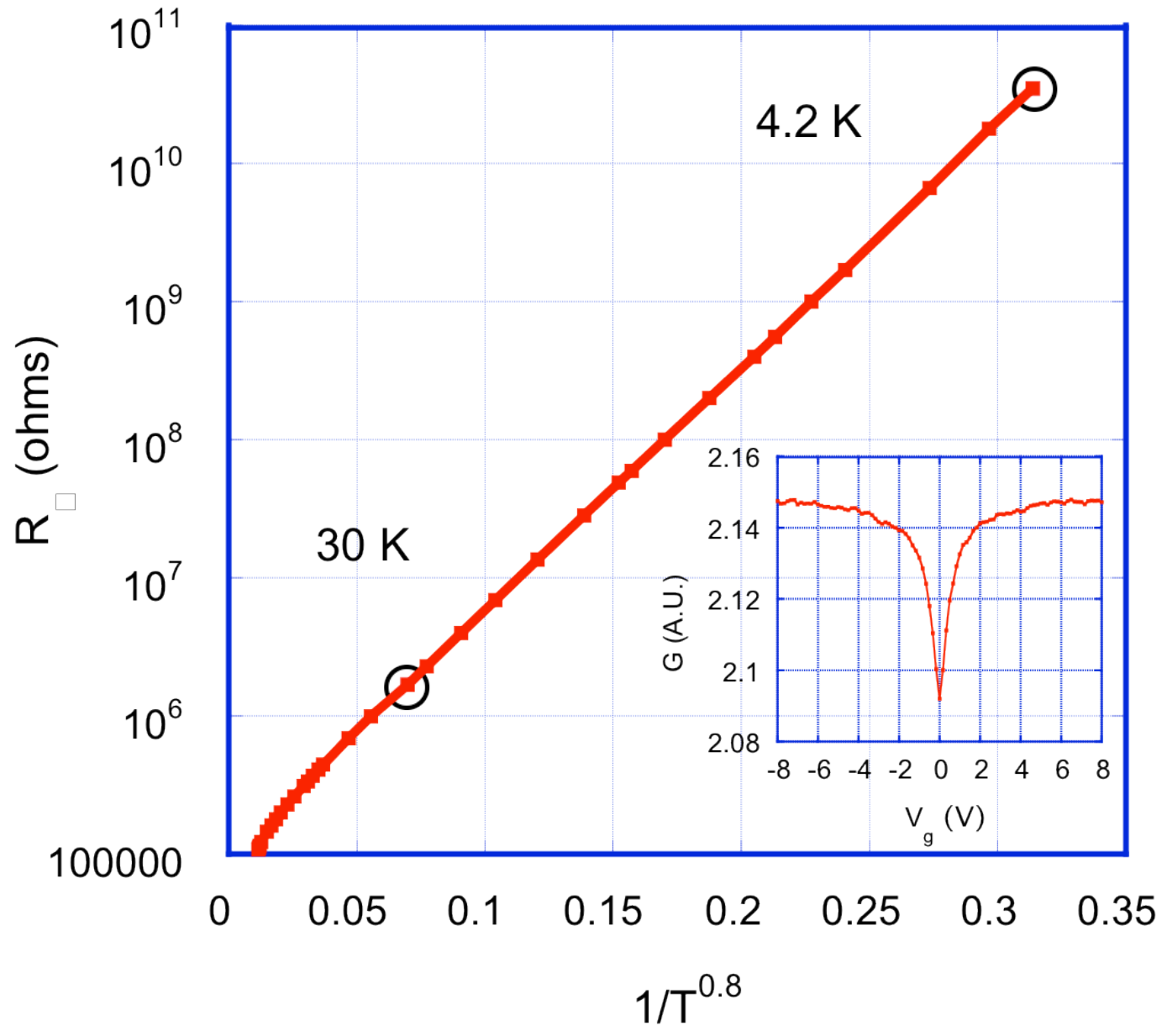

**Figure 1**: Resistivity curve of a 200 angstroms thick granular Al film used for the measurements and analyses shown in Figure 2 to Figure 7. Inset: conductance versus gate voltage curve showing a typical memory dip grown at 4.2K under $V_g$=0V.

## 3. Experimental results

### *3.1 Freezing of the memory dip upon cooling*

In the first series of experiments, we build the $V_{g\ build}$ memory dip at various $T_{build}$ and erase it at $T_{erase}$=4K, the protocol being repeated for several values of $T_{build}>T_{erase}$. We wish that the building and erasure steps be performed at well-defined constant temperatures, so whenever the temperature is being changed $V_g$ is set to $V_{g\ cool}$=-15V. Hence the sample is first cooled from high temperatures (at least 100K) to $T_{build}$ under $V_{g\ cool}$=-15V, then $V_g$ is switched to $V_{g\ build}$ =0V and a memory dip is built there for $t_{build}$=30 minutes. $V_g$ is then switched back to -15V and the sample is immediately quenched from $T_{build}$ to $T_{erase}$=4K by plunging it in the liquid helium bath. As soon as its temperature is stabilized at 4K, which is realized within 10-15 secs, $V_g$ is switched to $V_{g\ erase}$=+15V were a new memory dip is of course formed. From time to time a full $V_g$ scan is performed in order to measure the vanishing memory dip at $V_{g\ build}$ =0V as well as the increasing new one at +15V, the shape of which will be compared to the vanishing one. The sample itself is used as the thermometer so that its temperature is always precisely known.

For the results shown below we used a 200 angstroms thick insulating granular Al film, whose resistance versus $T$ curve is shown in Figure 1 together with a typical memory dip built at 4K and $V_g$= 0. Details on the samples fabrication and characterization were given in [7].

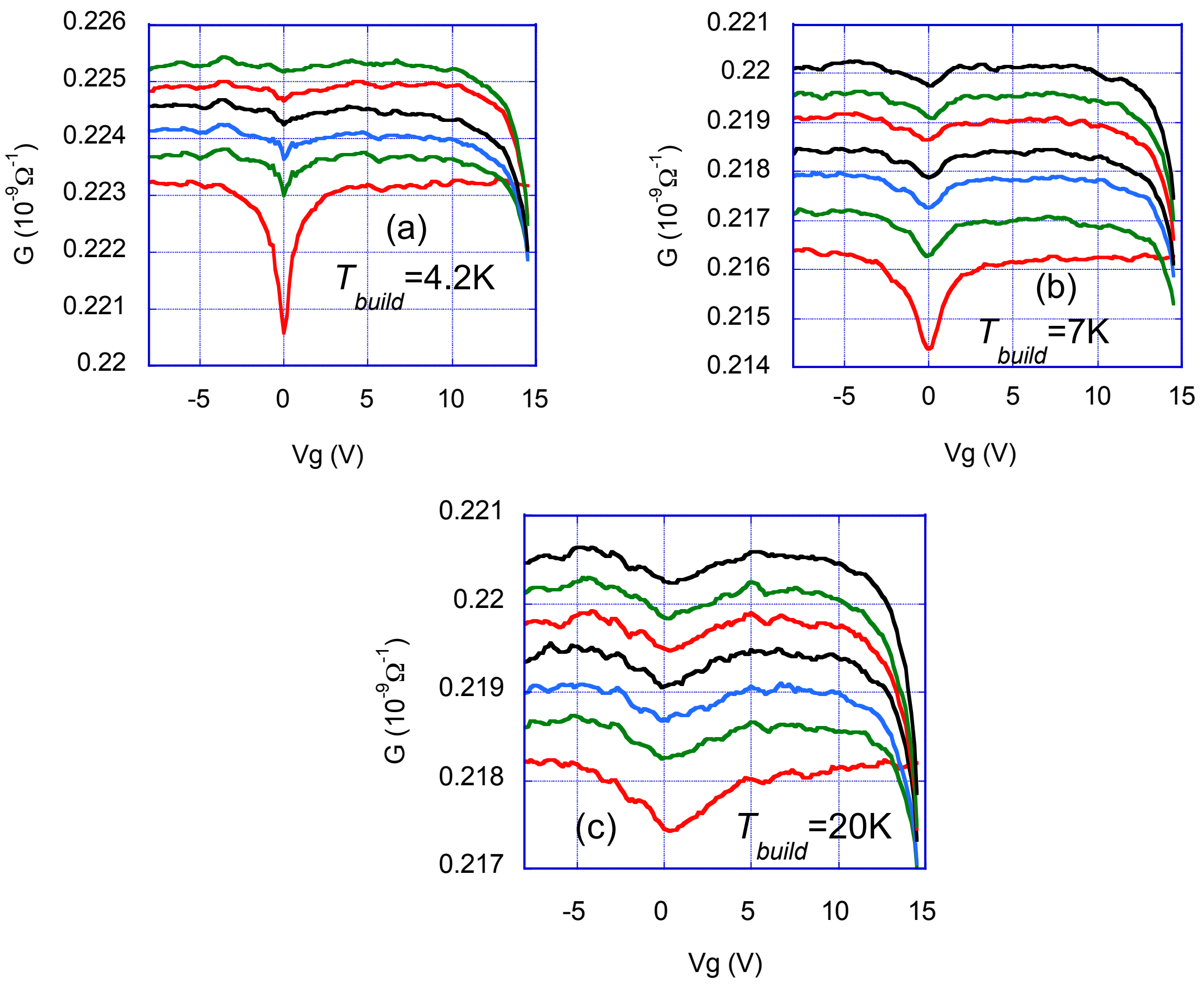

**Figure 2** : Time evolution of memory dips formed at various temperatures $T_{build}$ under $V_{g\ build}$=0V and during $t_{build}$=30 minutes, as they are erased after a cooldown to $T_{erase}$ =4.2K. In each set of curves time increases from bottom to top. The first scan is performed immediately after the cool down, successive scans are separated by 104 minutes time intervals. In (a) we show scans n° 1, 2, 3, 4, 5 and 8; in (b): n° 1, 2, 3, 4, 5, 6 and 8; in (c) n° 1, 2, 3, 4, 5, 8 and 10. One sees the progressive erasure of the memory dip at $V_{g\ build}$=0V and the narrow memory dip growing at $V_g$=+15V. Curves are shifted for clarity.

In Figure 2 we show for three $T_{build}$ values, how the $G(V_g)$ curves evolve with time after $V_g$ has been set to +15V. In each set of curves the lowest trace is the first one measured at $T_{erase}$ showing the memory dip just before the erasure starts, the higher ones are subsequent traces. Previous studies have shown [7] that although the shape of isothermal memory dips does not vary much with $T$ except below ~10K, their width (FWHM) does vary significantly and is proportional to $k_BT$. It is seen in Figure 2 that the width of memory dips is frozen upon cooling : they all retain their original $T_{build}$ width even after a long time spent at $T_{erase}<T_{build}$. In other words there is no thermalization of memory dips at low temperature. Note that if the gate voltage was kept at $V_{g\ build}$ =0V this freezing of the width would be obscured by the overgrowth of an ample new narrow memory dip, which in our protocol is observed at $V_{g\ erase}$ =+15V. As for the amplitudes one clearly sees that the erasure dynamics strongly depends on the difference between $T_{build}$ and $T_{erase}$. The highest traces correspond to erasure times of $30t_{build}$. In the isothermal case after such a time the initial memory dip has been essentially erased, while for the MD quenched from 20K to 4.2K only half of its amplitude has disappeared. It is thus evident that the erasure of the memory dip is much slower when $T_{erase} < T_{build}$ than in the isothermal case. This conclusion is drawn from the direct observation of raw data, and does not involve any comparison or normalization using two different memory dips. Thus the problems associated to such manipulations (see Appendix A) are avoided.

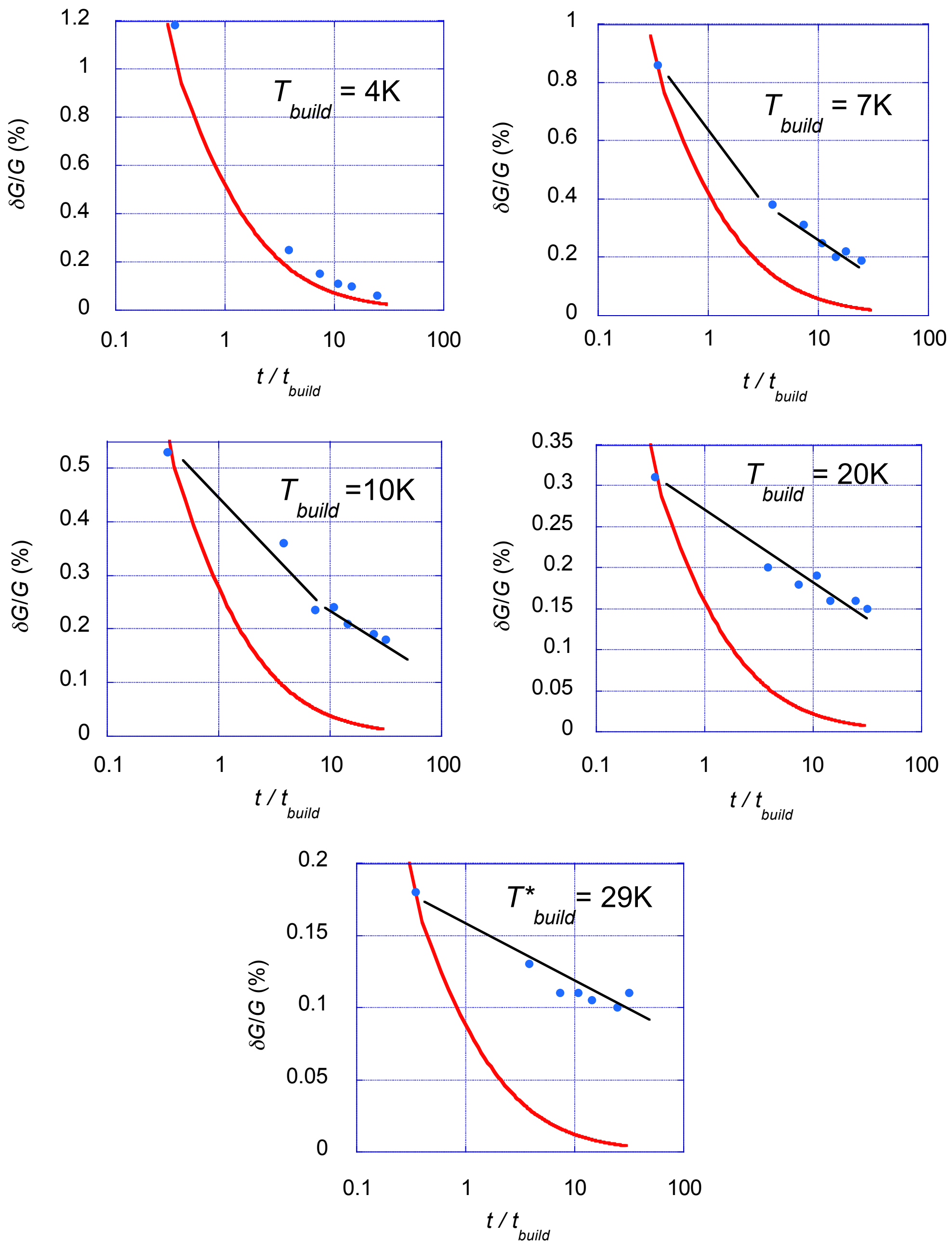

**Figure 3** : Time evolution of the memory dip amplitude during its erasure at $T_{erase}$=4.2K, for various $T_{build}$>$T_{erase}$. Amplitudes are given in % of the (4.2K) conductance. The continuous curves show the expected isothermal behaviour. The differences with the experimental points demonstrate the freezing of the dynamics.

In Figure 3 we plotted the time evolution of the fading memory dip amplitude, for various $T_{build}$ values. It is seen that when $T_{build}$=$T_{erase}$=4K (isothermal experiment) the amplitude approximately follows Eq 1 shown as a continuous line on the graphs. However it is clear that when $T_{build} > T_{erase}$=4K, such is not the case and it takes a much longer time to erase the memory dip, an effect all the more pronounced as $T_{build}$ is higher. From these data the $T_{build}$ dependence of the characteristic erasure time is clear, and it is seen that a memory dip formed at $T_{build}$>>4K is substantially frozen at 4K.

### 3.2 Accelerated erasure of the memory dip upon heating

Conversely the activation of the dynamics should manifest itself as an acceleration of the erasure and thus a decrease of the erasure time when $T_{erase}>T_{build}$. In Figure 4 left we show the results of a series of measurements with $T_{build}$=4K and $T_{erase}>T_{build}$. As an acceleration of the dynamics is looked for, smaller $t/t_{build}$ values than in Figure 3 are investigated. The amplitude of the memory dip is seen to depend significantly on $T_{erase}$ but what we are interested in is the time it takes to erase the dips. Unlike in Figure 3 it is not obvious that the curves extrapolate to zero at significantly different times. As seen in Figure 4 right, when normalized, all the curves including a 4K isothermal one are superposed within the experimental error. This extends to a higher $T_{erase}$ range, and corroborates, our previous result of [7]. But while it was thought previously to indicate the absence of thermal activation, we now know that it cannot be correct since it would otherwise contradict the results of Figure 3.

The simplest interpretation of this apparent paradox is the following. When T is increased from $T_{build}$ to $T_{erase}$ the relaxation times that were involved in building the memory dip are shifted to quite short times, such that only parts of the long time tail of the erasure curves can be probed. The shorter time logarithmic parts, which give an easy estimate of the erasure time by extrapolation to zero, cannot be observed. Unfortunately the long time tails have more or less the same shape and when normalized they are roughly superposed. So our measurements are not accurate enough to extract the erasure times from them. This argument is quantified in the next section.

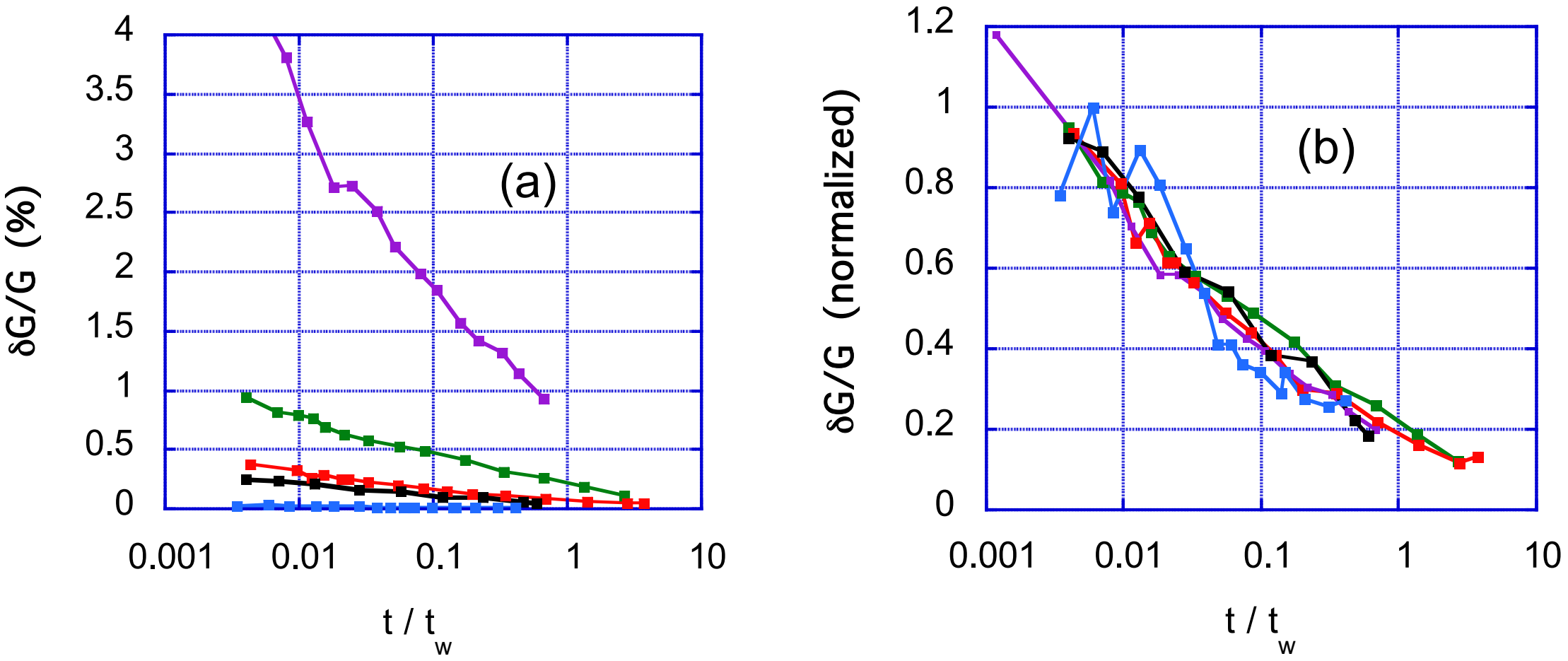

**Figure 4** : Time evolution of the memory dip amplitude during its erasure at various $T_{erase}$ ($T_{build}$=4.2K). In (a) amplitudes are given in % of the conductance. From bottom to top $T_{erase}$=20K, 12K, 10K, 8K, 4.2K. In (b) on sees that the normalized curves are superposed within the experimental scatter and extrapolate to zero at the same time. No indication of an accelerated dynamics when $T_{erase}$ >$T_{build}$ appears, in apparent contradiction with Figure 3.

The acceleration of the dynamics can nevertheless be demonstrated in the following way. Instead of performing the whole erasure at $T_{erase}$ larger than $T_{build}$, we perform it at the same temperature as $T_{build}$ (4K), *except for a short excursion to a larger* $T_{erase}$. The amplitudes of the fading memory dip measured before and after the higher $T$ excursion can be compared without any normalisation as they are all measured at the same $T$. The acceleration of the dynamics should manifest itself as a discontinuity in the curves.

In practice a memory dip was formed during 1 hour at 4.2 K and $V_{g\ build}$ = 0V, then $V_g$ was set to +20V, scans being performed from time to time to measure the fading memory dip amplitude at 0V. The temperature excursion was performed between the first and second scans: $T$ was raised (raising time 30-45 seconds) and maintained at a higher value for one minute and then set back to 4.2K (cooling time around 15 seconds). Figure 5 shows the erasure curves obtained for various excursion temperature values. It is seen that a discontinuity is induced in the curves, which is larger the higher the excursion temperature. This reveals the significant acceleration of the erasure dynamics during the higher temperature excursion. Note the important point that the memory dip measured at 4K after the high temperature excursion has its normal 4K width. This is seen in Figure 6 and shows that the much smaller amplitudes measured after the discontinuity are due to an accelerated erasure and not to a thermal broadening caused by the higher temperature excursion.

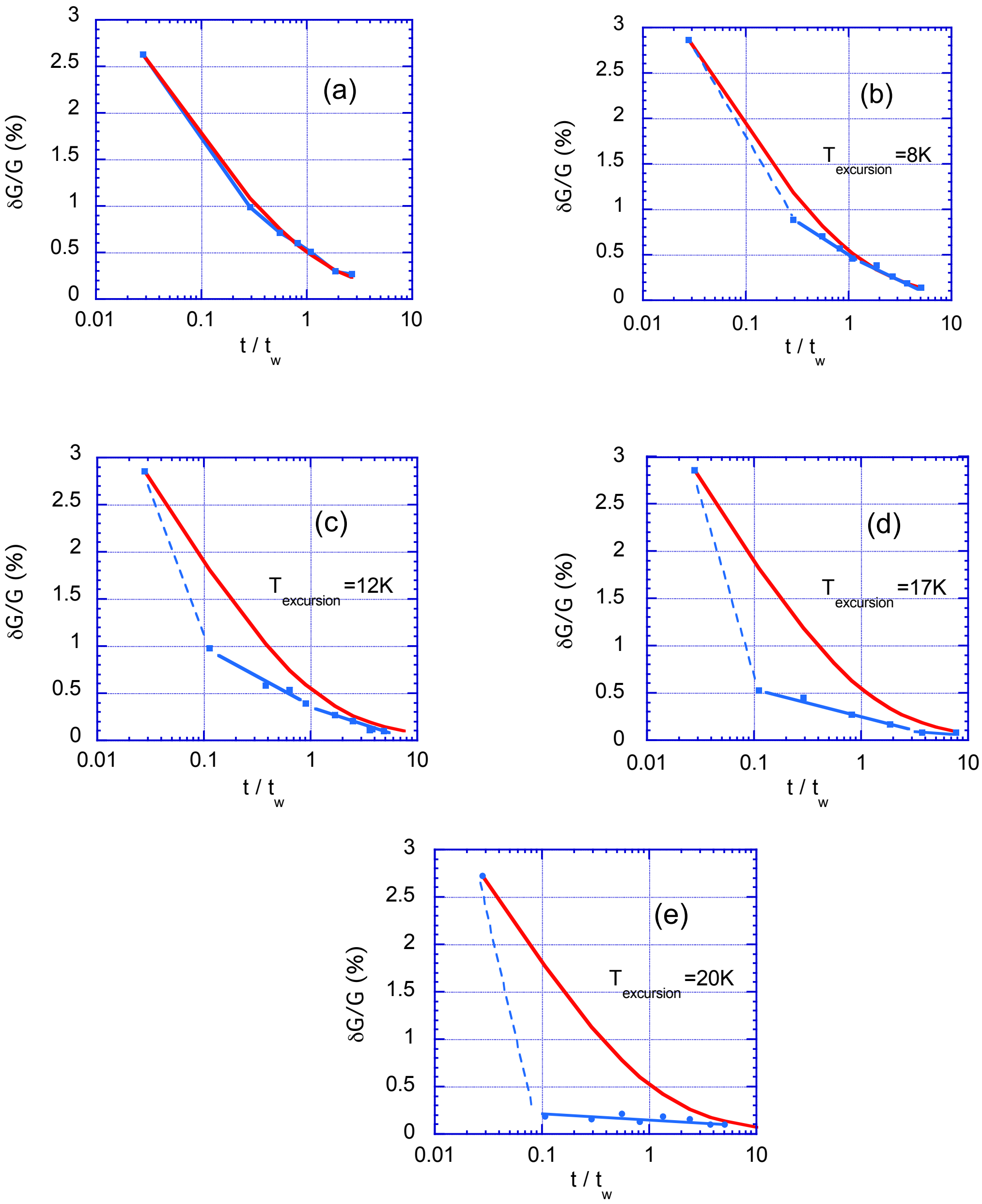

**Figure 5**: Time dependence of memory dips formed at $T_{build}$=4.2K during 1 hour and erased at $T_{erase}$= 4.2K except for a short excursion at a higher temperature $T_{excursion}$ indicated on the graphs (temperature rise duration : 30-45 secs, stay at $T_{excursion}$ : 60 secs, cooling back to 4.2K within around 15 secs). The discontinuity created in the curves demonstrates the acceleration of the erasure during the short high T excursion.

The data presented above convincingly demonstrate that the slow conductance relaxations are thermally activated and that the dynamics is not dominated by elastic tunneling events.

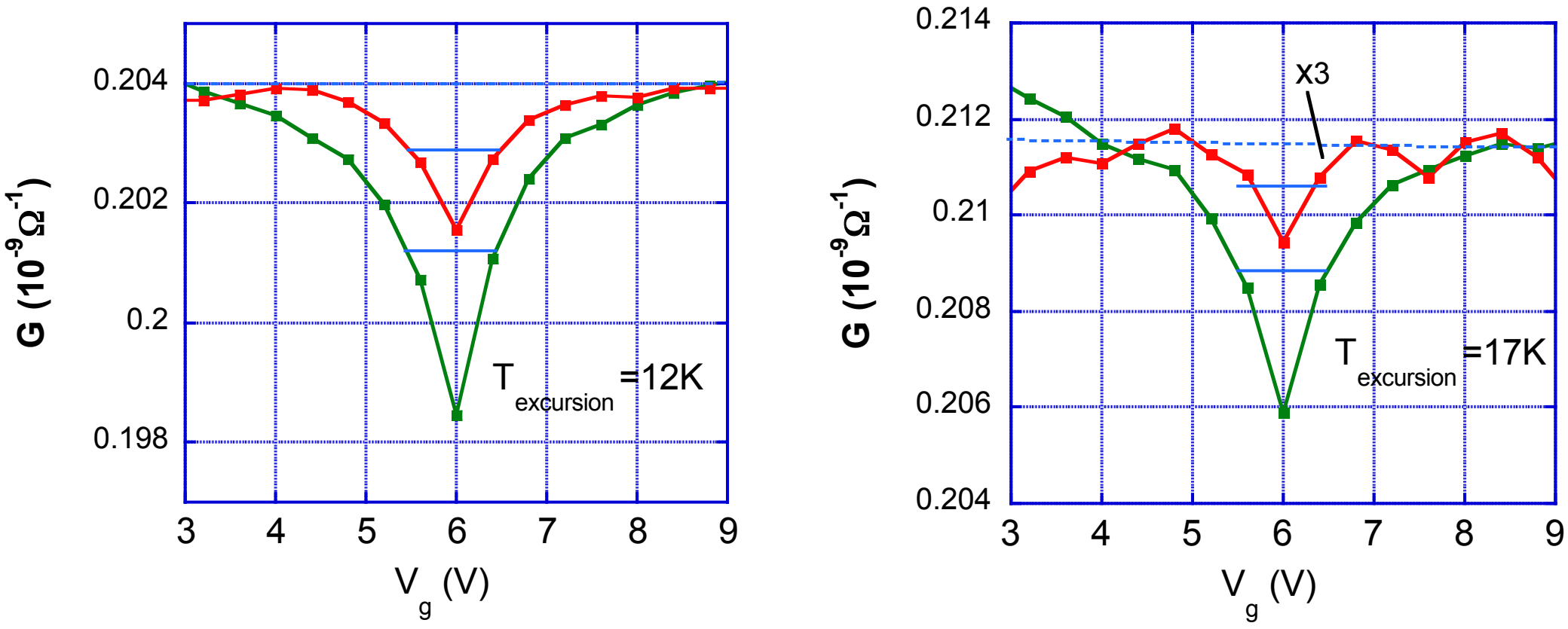

**Figure 6** : Examples of memory dips measured just before and just after the higher temperature excursion (see Figure 5). Left: $T_{excursion}$=12K, right: $T_{excursion}$=17K. For $T_{excursion}$=17K the amplitude of the dip measured after the excursion is multiplied by 3 to be more visible. It is seen that while the amplitude is substantially reduced by the excursion, the width is unchanged and acquires no thermal broadening.

### 3.3 Activation energy range

We may try to fit the curves of Figure 3 to get an estimate of the activation energies involved. Following [7] the memory dip amplitude being erased is modeled by :

$$\delta G(t) = \delta G_0 \sum_i \left[1 - \exp\left(-\frac{t_w}{\tau_i}\right)\right] \exp\left(-\frac{t}{\tau_i}\right) \qquad (2)$$

where $t$ is measured starting from $t_w$ and the $\tau_i$'s are the relaxation times of the slow modes involved in the memory dip formation. If $\tau_i = \tau_0 \exp(\xi_i)$ and the $\xi_i$'s have a broad flat distribution then equation (1) is obtained. Inserting activation energies we may more generally write :

$$\tau_{ij}(T) = \tau_0 \exp(\xi_i) \exp\left(\frac{\Delta E_j}{k_B T}\right) \qquad (3)$$

Choosing a single $\xi_0$ and broadly distributed $\Delta E_j$'s implies that during the building of the memory dip, an activation energy interval essentially determined by $T_{build}$ is selected. It turns out that this does not permit to fit the whole set of erasure curves properly. A broad distribution of $\xi_i$'s is necessary and the whole set of curves of Figure 3 can be fitted quite well with a single activation energy $\Delta E_0 \approx 3 \pm 0.5 meV$ . The fit is a bit better if we also take a distribution of activation energies $\Delta E_j \in [0, \Delta E_{\max}]$, $\Delta E_{max}$ being the fitting parameter. We assume that the $\xi_i$'s and $\Delta E_j$'s are not correlated (see Appendix B for a more systematic discussion). In Figure 6 we show the result of the quite satisfactory fit from which we get $\Delta E_{\max} \approx 5 \pm 0.5 meV$ .

Concerning the analogous experiment with $T_{erase}$ > $T_{build}$ (Figure 4), the similar numerical calculation with $\Delta E_{max}$=5meV yields the curves shown in Figure 8. One sees that in the whole temperature range a small change should be observed between the 4K isothermal curve and the one with $T_{erase}$=29K. But the expected difference is of the same order as the experimental points scatter, which may explain why we cannot see it in the data, as anticipated in the previous section.

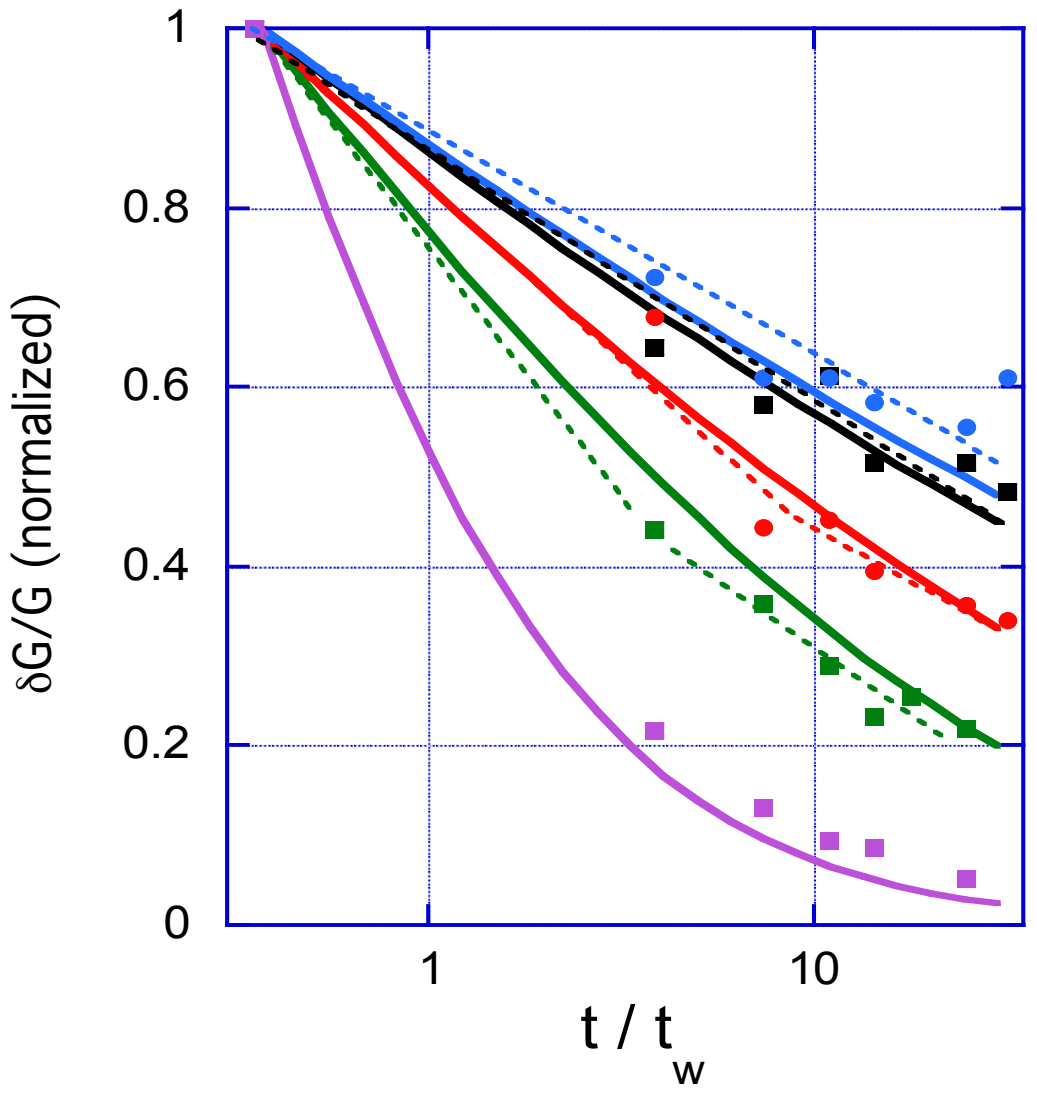

**Figure 7:** Normalized erasure curves in the case $T_{erase}< T_{build}$ ($T_{erase}$=4.2K). The experimental points are those of Figure 3, the dotted lines are guidelines through the points. The continuous lines are calculated using a flat distribution of activation energies [0, $\Delta E_{max}$]. The fitting parameter $\Delta E_{\max} \approx 5 \pm 0.5 meV$ . See text for more details.

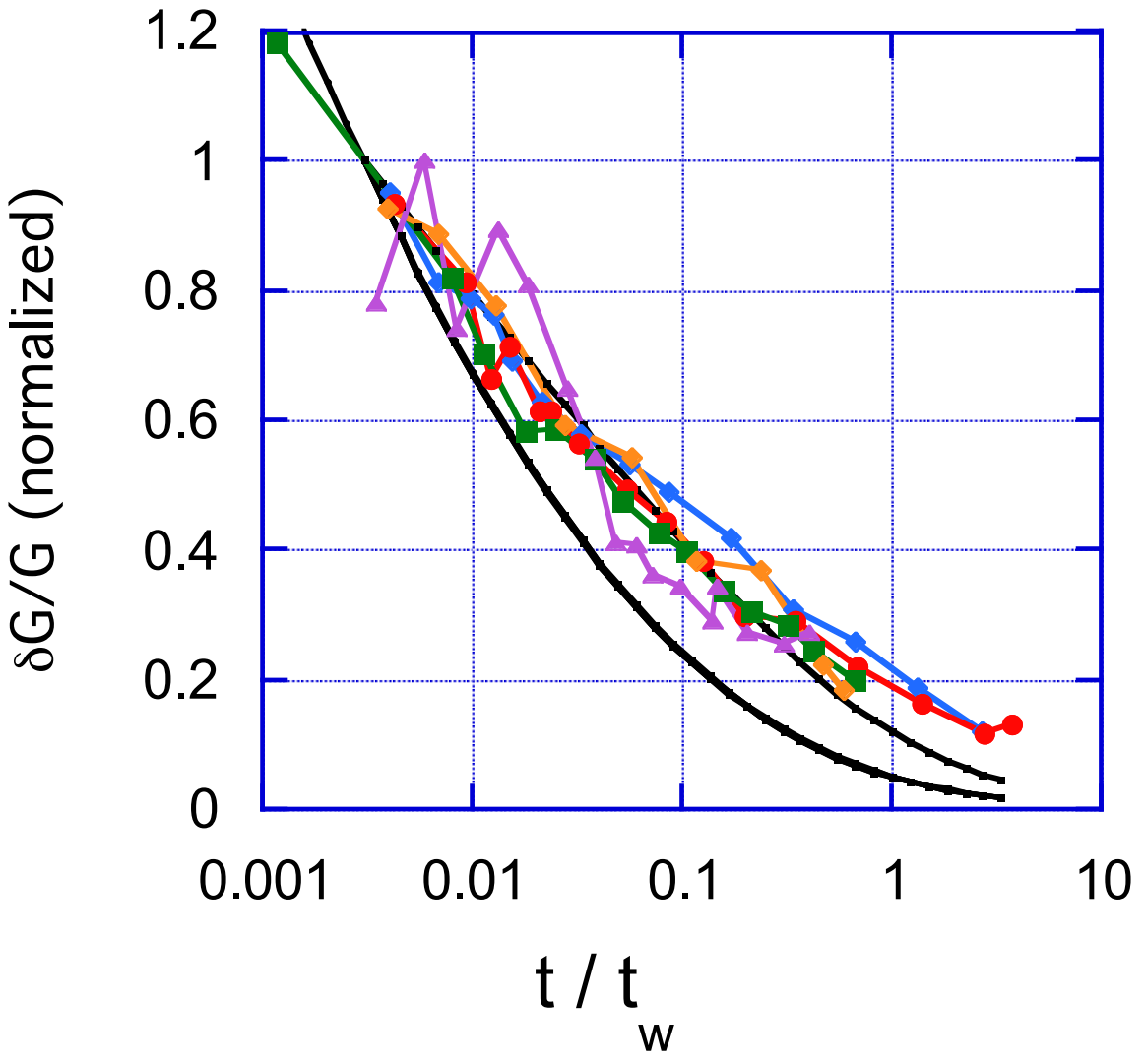

**Figure 8**: Normalized erasure curves in the case $T_{erase}>T_{build}$ ($T_{build}$=4.2K). Experimental points are those of Figure 4. The smooth lines are two curves calculated for $T_{erase}$=4K and $T_{erase}$=29K and using the same activation energy distribution as in Figure 7. One sees that the difference between the extreme curves are not expected to be larger than the experimental scatter.

## 4. Discussion

*4.1 Activation law and activation energies*

We saw that the temperature variation of the slow dynamics is essentially of the Arrhenius type i.e. one can fit the data over the experimental temperature range with a fixed activation energy distribution. However in the framework of the electron glass ideas one does not necessarily expect such a dependence. Like for variable range hopping conductance, the optimization of tunneling probabilities for multi-electron processes may produce a non-Arrhenius behaviour with typical activation energies increasing with the temperature. Our present data do not give evidence of this, but we note that the temperature dependence of our samples conductance is also closer to Arrhenius than to any standard variable range hopping law. It would be interesting to explore broader resistance and temperature ranges as well as systems having Mott or Efros-Shklovskii variable range hopping conductances, to see whether the temperature dependence of the dynamics is different.

As for the presently studied system we may like to compare the activation energy range of the dynamics with typical electronic energy scales, and see whether they are correlated. The typical Al grain size of our samples is 2-5 nm. Assuming 1 nm thick alumina layers separating the grains, we can estimate the average charging energy of a single grain inside the material as $E_c \approx 20meV$ . However the energy scale determining the conductance is smaller. As seen in Figure 1 a quite good empirical fit of $\sigma(T)$ in the temperature range of interest is :

$$\sigma = \sigma_0 \exp\left[-\left(\frac{T_0}{T}\right)^{0.8}\right] \qquad (4)$$

with $T_0 \approx 100K \equiv 8.7meV$ . We shall not discuss here the precise meaning of the $T$ dependence of the conductance and of $T_0$. One may also approximately fit the low temperature part with an Arrhenius law and get $T_{0(Arrhenius)} \approx 50K \equiv 4.4meV$ . Thus the energy scale one may extract form the conductance is of the same order as $\Delta E_{max}$.

To see whether they are correlated, we performed the same dynamical experiment on a less insulating sample obeying Eq. (4) with a significantly smaller $T_0 \approx 43K \equiv 3.7meV$ . Its erasure curves at 4K for dips built at various $T_{buid}$>4K are shown in Figure 9. It is seen that the curve with $T_{build}$=8K is well fitted with $\Delta E_{max}$=5meV, exactly like with the more insulating sample. The other curves with larger $T_{build}$ seem to need a somewhat higher $\Delta E_{max}$, which whithin our simple description may indicate a narrower distribution of $\xi_i$'s than in the more insulating samples, but the scatter of the points is larger so that we will not elaborate too much on this.

Comparing with the previous sample, the conductance and slow dynamics activation energies do not seem to be directly correlated: dividing $T_0$ by a factor of two does not seem to alter $\Delta E_{max}$ the same way. This may mean that the slowly relaxing modes and the much faster electronic processes involved in the conductance itself, are to some extent independent. It may be that they are of different nature, or that the samples being strongly heterogeneous, the slow electronic processes occur in highly insulating areas which the current carrying path skirt rather than cross.

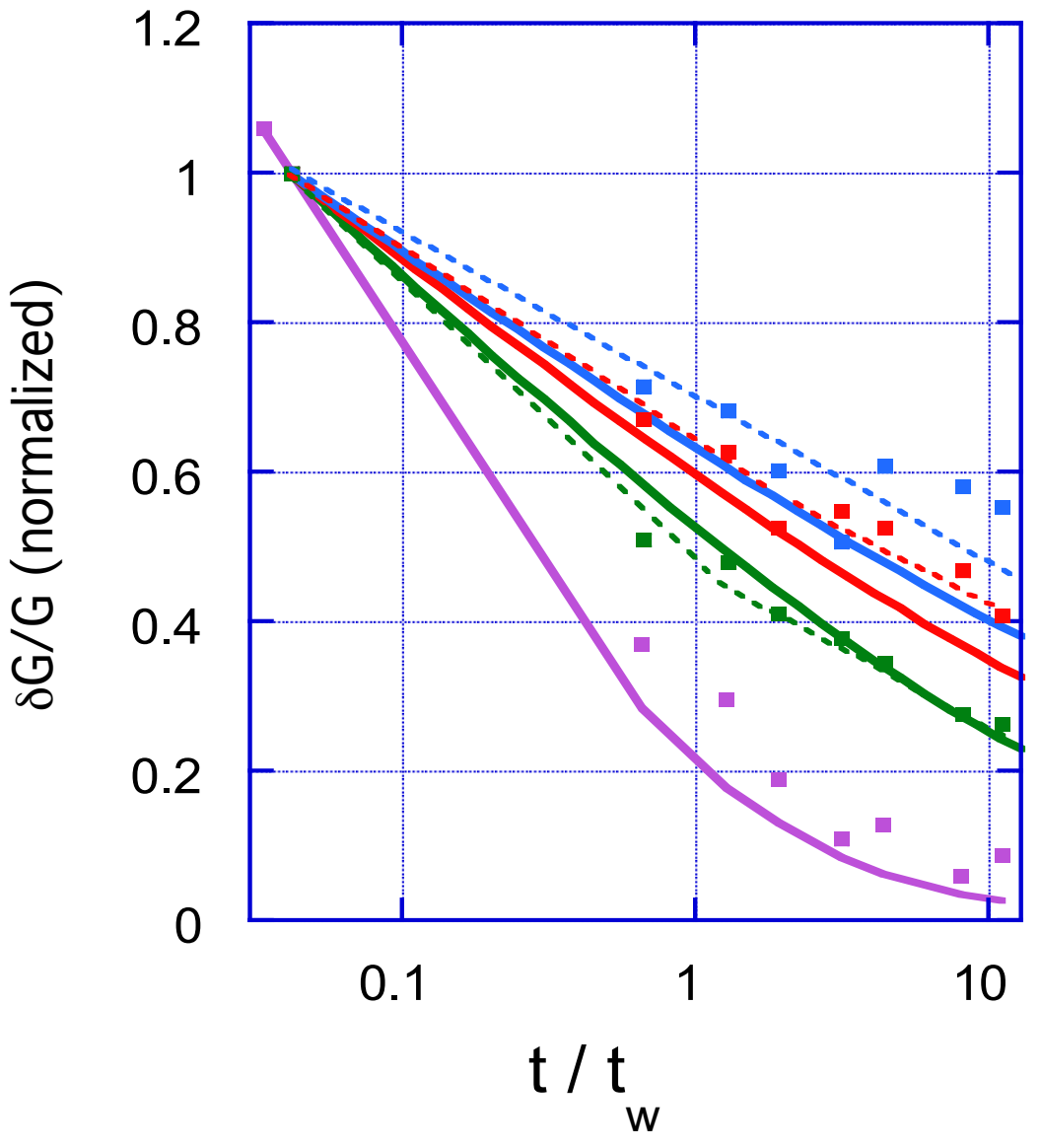

Figure 9: Normalized erasure curves in the case $T_{erase}$>$T_{build}$ ($T_{erase}$=4.2K) for a more conducting sample than in Figure 7 ( $R_{\square} \approx 1.5 10^7 \Omega$ ). Like previously the dots are the experimental points, the dotted lines are guidelines through them and the continuous lines are calculated using the same distribution of activation energy with $\Delta E_{\max} \approx 5meV$ .

Interestingly the observation of a freezing of the dynamics puts the question of any finite temperature glass transition in a new experimental perspective. Such a finite temperature phase transition has been predicted in a mean field treatment of the electron glass [9], a transition that hasn't been observed yet in numerical simulations of electron glass models containing site energy disorder [1]. As far as our experiments with granular Al films are concerned, only a gradual freezing is observed, with no signature of any well-defined finite temperature transition in the temperature range explored.

The questions raised by the precise $T$ dependence of the dynamics, and the link between slow relaxation and conductance activation energies need further investigations in a broader resistivity and temperature range as well as in other systems.

*4.2 Effects of the high temperature history*

Due to freezing, traces of the high temperature gate voltage history of a sample are present at low temperature, and may be impossible to eliminate without heating. This can explain previous observations that did not have a clear interpretation. The freezing of a thermally broadened memory dip is the cause of the "thermal memory" mentioned previously [7], and the evolution of a memory dip shape after it has been cooled ($V_g$ kept inside it), corresponds to the overgrowth of a narrower part. Accordingly the shapes of memory dips built during cool-downs are different from isothermal ones, as shown in Figure 10 below. We show the superposition of a memory dip obtained after a quench from 67K and a further growth for a few hours at 4K, with an isothermal 4K one. The frozen modes contribute significantly to the overall amplitude of the first memory dip and broaden its feet. We illustrate it in the case of a highly insulating sample ( $R_\square \approx 5.10^{10}\,\Omega$ ). In the less insulating and more commonly studied samples, the effect is less prominent: the lower the resistance per square, the smaller the amplitude of the frozen high T contributions as compared to the low temperature ones.

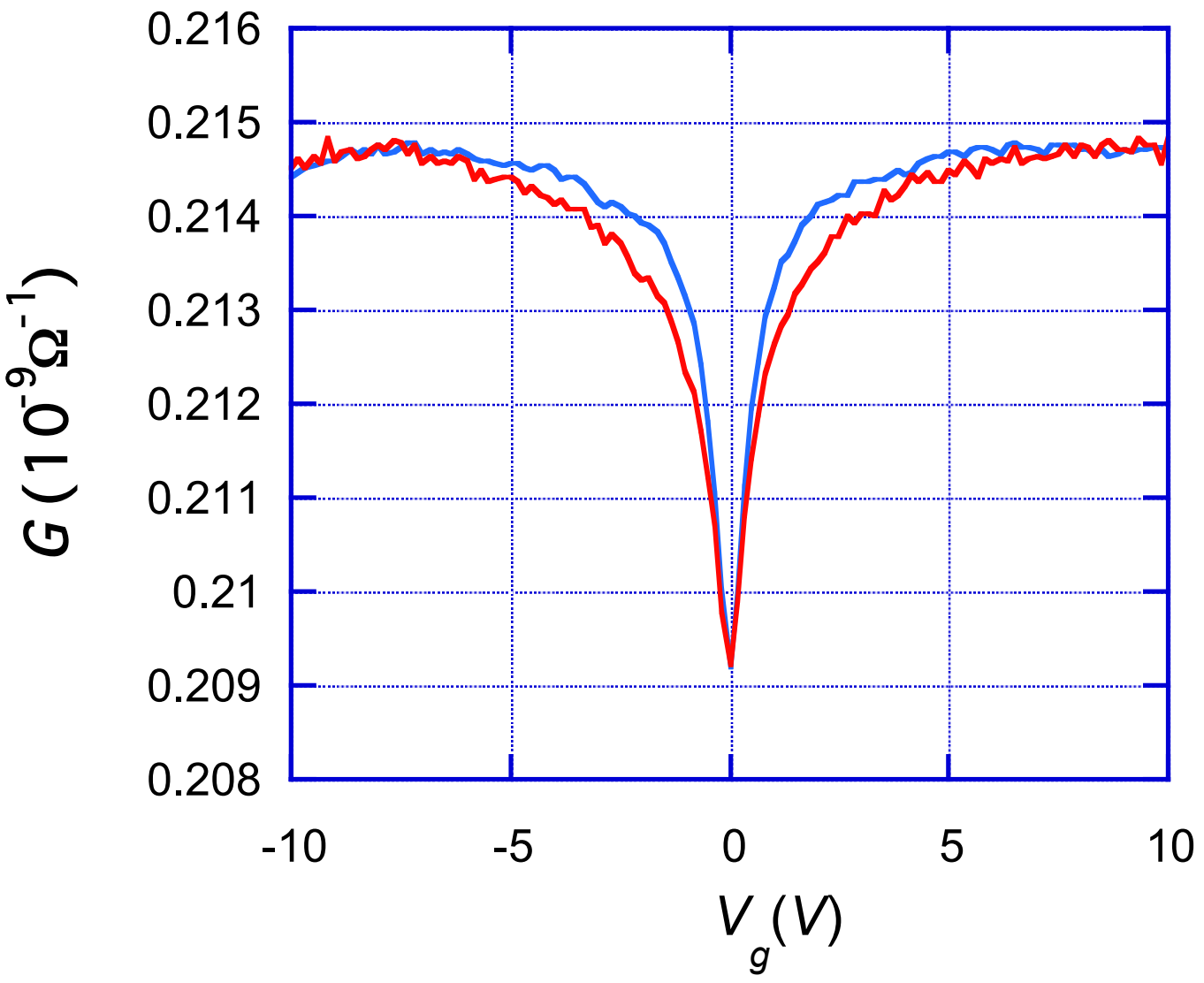

**Figure 10** : Comparison of the shapes of two (normalized) memory dips measured at 4K on the same sample. The thinner dip is an isothermal one entirely grown at 4K. The broader one was grown after a quench cool from 67K and subsequent grown for a few hours at 4K, under the same gate voltage. The difference arises from a broader frozen contribution accumulated during the cooling.

The relative amplitudes of frozen and isothermal low temperature contributions vary a lot among systems. Thus in NbSi [4] one can observe very broad and intense frozen contributions that may make it difficult to isolate and study the low temperature parts. As we already mentioned, in discontinuous gold [3] the low temperature dips are much smaller than the high temperature ones. In the opposite, some indium oxide samples seem to have a very rapidly decreasing memory dip amplitude as the temperature is increased from the liquid helium range [10]. In this case frozen contributions are expected to be small and not to disturb low $T$ measurements. But conversely this makes thermal activation more difficult to detect : erasure measurements should be performed in a lower $T$ range than in the present study.

*4.3 Case of indium oxide*

The present results on granular aluminum call for a re-examination of the case of indium oxide. In [11] the question of the temperature dependence of the slow dynamic was considered in the low temperature range ($T$<6K). The use of a "double conductance excitation" protocol led to the conclusion that highly doped $InO_x$ have a $T$ independent dynamics, while in lightly doped $In_2O_{3-x}$ the dynamics accelerates when $T$ is reduced. However as these experiments are isothermal, it is not clear to us how they can reflect the $T$ dependence of the dynamics. A more detailed discussion of this point is given in Appendix A. Preliminary results of a similar protocol applied to granular Al show no $T$ dependence of the "viscosity parameter" defined in [11]. This quantity is thus not sensitive to thermal activation, which puts into question conclusions drawn from it.

As far as we know the erasure protocols used in the present work have not been applied to indium oxide. However the memory of high temperature thermal width was shown in [12] : in Figure 10(c) a memory dip built at $T_{build}$=4.1K is shown to retain its initial width after 8 mins at 1.6K. This is consistent with, but not yet a proof of, thermal activation and freezing. A measurement of the erasure time of this dip would give a definite answer. We feel that as of today the question whether the slow dynamics in indium oxide is or isn't activated and the glass at hand is classical or quantum, is still open.

## 5. Conclusion

We have shown that the slow dynamics observed in insulating granular aluminum thin films is thermally activated, unlike what was previously believed. For the samples studied, the behaviour is close to Arrhenius with an activation energy of the order of 30K. In the temperature range studied (4K-30K) no signature of a glass transition temperature is observed. We showed why previous experiments missed the temperature dependence of the slow dynamics and led to the wrong conclusion that it is solely due to quantum tunneling.

The simple erasure protocol we used in the present study does not rely on any normalization procedure or comparison of different memory dips. It is thus free of the uncontrolled assumptions present in other used protocols. We suggest that it is used in all the other systems so that reliable results are obtained and sound comparisons can be made.

In particular we argued that previous experiments on indium oxide are not suitable to test the temperature dependence of the dynamics. It may well be that by using our protocol, thermal activation is found in this system too, which would dismiss the quantum glass hypothesis. Thermal activation would then appear to be a universal feature in all Anderson insulators showing slow conductance relaxations associated to memory dips.

The determination of activation energies provides a new grasp on the physics at work. Its systematic study in different systems, and comparisons with specific predictions of the theoretical and numerical approaches of the electron glass [1], which are all in the classical limit, may bring new knowledge about the mechanisms of glassiness observed in disordered insulators.

**Acknowledgement**

We acknowledge several fruitful discussions with A. Frydman and M. Pollak.

## Appendix A: analysis of previously used protocols

For completeness and clarity, we discuss here briefly the various protocols used so far in the literature to study the temperature dependence of the slow dynamics in Anderson insulators, an their shortcomings.

*A.1 Relaxation slopes*

The mere existence of slow Ln($t$) conductance relaxations in a broad temperature range (from 1K to 50K) is not a proof that the dynamics is not thermally activated, but only that long enough relaxation times survive up to the high temperature limit, with a distribution function proportional to 1/τ. One cannot compare the dynamics of different samples or of the same sample at different temperatures by simply comparing their relaxation slopes, either $\frac{dG}{dLn(t)}$ or $\frac{1}{G}\frac{dG}{dLn(t)}$. This is because one does not know how to relate the amplitudes of the conductance changes to the advancement of the systems relaxations toward equilibrium. In most systems studied $G$ vanishes rapidly when $T$ is lowered at low temperature, so does also $\left|\frac{dG}{dLn(t)}\right|$. It seems more sensible to consider $\left|\frac{1}{G}\frac{dG}{dLn(t)}\right|$. This quantity increases significantly when $T$ is lowered, but nobody believes that this reflects a sharp acceleration of the dynamics upon cooling. Remarkably discontinuous metal films show an opposite behaviour [3], which is thought to demonstrate a freezing of the dynamics, but the erasure protocol used in this paper should be used to confirm and quantify this conclusion.

*A.2 Isothermal "two dip experiment" (TDE) and its variants*

A "two dip experiment" (TDE) protocol was developed in [13] to study doping effects in indium oxide films. In this experiment one first builds a memory dip for a given time $t_1$ (typically 24 hours), after which one shifts the gate voltage to another value thus starting to build a second memory dip. One then determines the time $\tau_{TDE}$ at which the amplitudes of the two dips become equal. We discussed this protocol in [6] in some detail. If the amplitudes of the two memory dips evolve logarithmically with time with equal slopes, like in granular Aluminum and highly doped $InO_x$, it is not expected to give any valuable information on the system dynamics as one simply expects $\tau_{TDE} \approx \sqrt{t_1 t_{scan}}$ (where $t_{scan}$ is the $V_g$ scan time used in the measurement). This dependence was checked experimentally with granular aluminum.

In [14] a "double conductance excitation " (DCE) protocol was discussed, which in our view is a mere variant of the TDE. In this case one builds the second memory dip for a fixed time $t_w$ and one computes the quantity $\eta = \frac{\Delta G_0}{\Delta G(t_w)}$ where $\Delta G_0$ is the initial amplitude of the first memory dip and $\Delta G(t_w)$ is the amount by which it was erased during $t_w$. If all dips amplitudes evolve like Ln($t$) with equal slopes then $\Delta G(t_w)$ is also the amplitude of the second memory dip after $t_w$. Again η does not contain any useful piece of information as it is expected to be $\eta = \frac{Ln\left(t_1/t_{scan}\right)}{Ln\left(t_w/t_{scan}\right)}$.

In [11] and [14] the TDE and DCE were used at various temperatures. For highly doped indium oxide they yielded no temperature dependence for $\tau_{\rm TDE}$ nor for $\eta$. This is quite natural owing to what we just said: by principle these protocols cannot reveal any temperature dependence on these samples. However the authors concluded that these results imply a temperature independent dynamics, a conclusion more or less explicitly based on the assumption that after $t_1$ the samples have reached full equilibrium. This assumption is unlikely and contradicts what is known about these systems, i.e. they never reach full equilibrium after a quench in any experimental time scale. The Ln($t$) relaxation may have flattened after $t_1$=24h, but it is not a constant and this make all the difference. Actually the highly doped indium oxide samples may have a thermally activated dynamics, this would not contradict the $T$ independent $\tau_{\rm TDE}$ and $\eta$. In our case we do not see any $T$ dependence of $\eta$ in highly insulating granular films, which glassy dynamics is thermally activated as shown in the present paper.

For less doped indium oxide samples, $\tau_{\rm TDE}$ and $\eta$ were shown to decrease when $T$ is decreased. This was interpreted as an acceleration of the dynamics as $T$ is reduced, a doubtful conclusion according to the preceding considerations. Several mechanisms may explain the results. The time evolutions of the two memory dips may not be symmetrical logarithmic relaxations [6], a fact which has actually been reported.

The results of the present paper also suggest a possible explanation. In the presence of activated dynamics, if the quench cool and growth of the first memory dip during $t_1$ are performed under the same $V_g$, then the dip contains a frozen contribution formed during the cooling. In granular aluminum this contribution is not a negligible part of the total amplitude even if the memory dip was grown for $t_1$=24 hours after the quench. It obviously alters the $\tau_{TDE}$ and $\eta$ values in a $T$ dependent manner, as its relative contribution to $\Delta G_0$ depends on the final temperature of the quench.

We conclude that isothermal experiments are not reliable to study any temperature dependence of the dynamics as long as relaxations are logarithmic. Protocols in which $T$ is changed are needed.

*A.3 Non isothermal "two dip experiment"*

In principle two dip experiments where the first and second memory dips are grown at different temperatures $T_1$ and $T_2$ may reveal the $T$ dependence of the dynamics.

In particular if the relaxation times have an Arrhenius dependence, their distribution function, and hence the slopes of the relaxation curves, is $T$ dependent. But as shown below in Appendix B it is not necessarily the case in the experimental time window where the Ln($t$) behaviour is observed. So non isothermal TDE is not necessarily sensitive to thermal effects. Moreover, the interpretation of the data relies on assumptions about the behaviour of the first memory dip amplitude when $T$ is changed from $T_1$ to $T_2$. In the case of granular aluminum, as seen in our present study, a memory dip grown at low $T$ has a much larger amplitude than another one grown at a higher temperature. So performing TDE with $T_1$>$T_2$ leads to the wrong conclusion that the dynamics is drastically accelerated when $T$ is lowered. This problem is only partially eliminated if, taking into account the thermal width of the memory dips, one uses their areas instead of their amplitudes to determine $\tau_{TDE}$.

Hence the quantitative analysis of non isothermal TDE experiments is not as straightforward as it would seem.

*A.4 "Erasure" protocol*

The erasure protocol described and used in the present study, including higher temperature excursions to probe the acceleration of dynamics upon heating (see main text), is free of the problems considered above, and is arguably the only reliable protocol used so far to test temperature effects on the slow dynamics. It demonstrated an activated kind of temperature dependence in amorphous NbSi and granular Al films.

**Appendix B: relaxation time distribution**

We discuss here the model used to fit the data of the present study. We suppose that the relaxation time of each relaxing "mode" is of the form

$$\tau_{ij} = \tau_0 \exp(\xi_i)\exp\left(\frac{\Delta E_j}{k_B T}\right) = \tau_0 \exp(m_i) \ ,$$

where due to disorder the parameters $\xi_i$ and $\Delta E_j$ are uncorrelated and broadly distributed, with constant distribution functions equal to $N_{\Delta E}$ and $N_\xi$ between zero and maximum bounds $\xi_{max}$ and $\Delta E_{max}$. One then obtains the distribution function of $m$ depicted below:

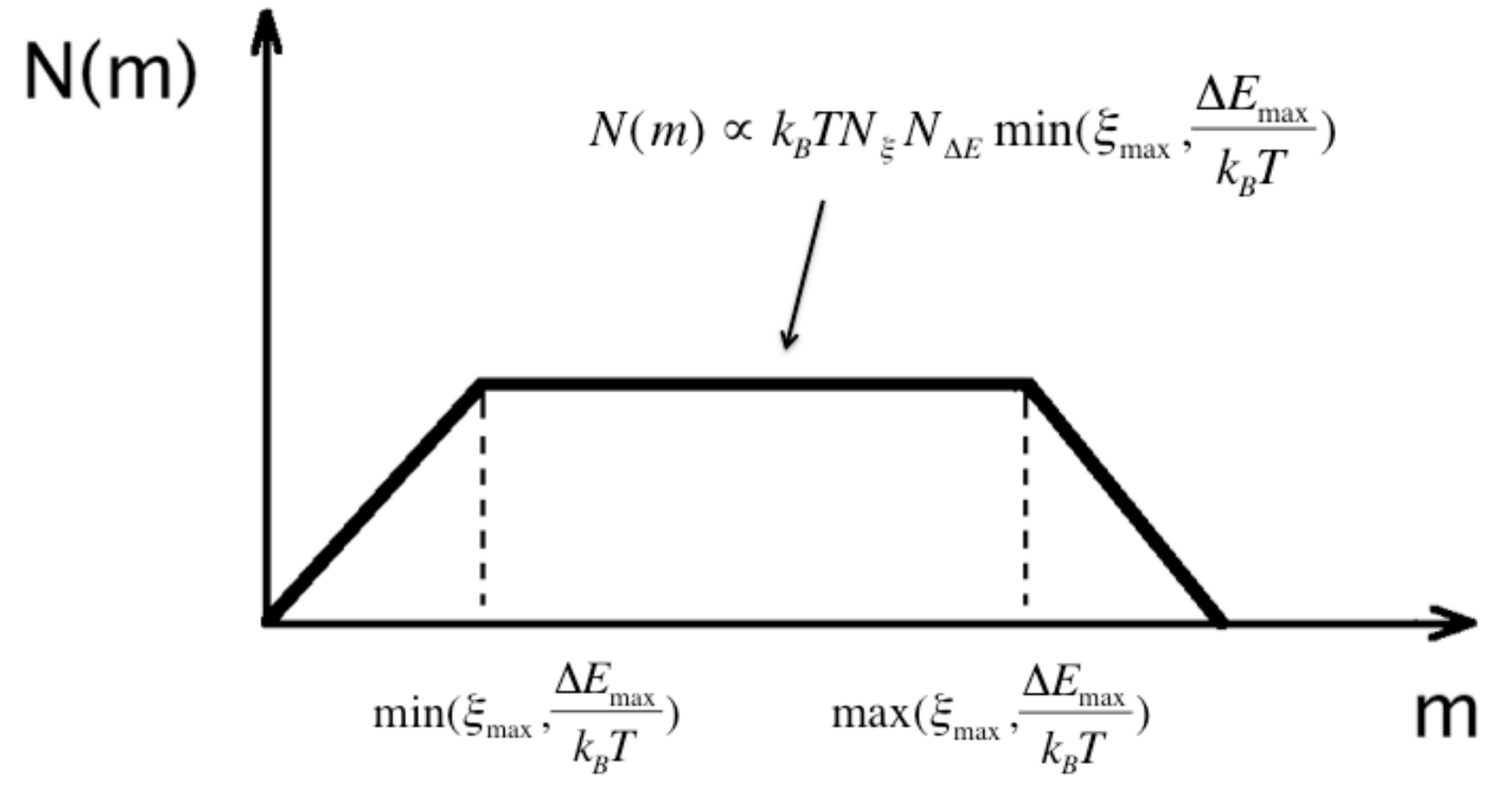

Since only logarithmic relaxations are experimentally observed, the practically accessible values of *m* lie in the flat part of the distribution. Depending on the relative values of $\xi_{max}$ and $\frac{\Delta E_{\max}}{k_B T}$ the flat part of the distribution may or may not depend on *T*. In Figure 7 it is seen that the temperature dependence of the curves saturates when $T_{build}$ is if the order 20-29K. This means that $\frac{\Delta E_{\max}}{k_B}$ is of the same order. Nevertheless slow relaxations still occur at these temperatures, which provides a lower bound for $\xi_{max}$. Relaxations go on for at least one hour at 30K, then with the typically used value $\tau_0 \approx 10^{-12}\,\mathrm{sec}$ one gets $\xi_{max} > 35$. Thus in the whole experimental temperature range one has $\xi_{\max} > \frac{\Delta E_{\max}}{k_B T}$ and the flat part of $N(m)$ is temperature independent. For all practical purposes we may take $\xi_{max}$ infinite. Thus for each value of $\Delta E_j$ one has a broad set of $\xi_i$, giving rise to a logarithmic contribution to the MD growth during the building step of duration $t_w$, and a contribution to the erasure curve of the form: $\delta G_j = \delta G_0 Ln\left[1+\frac{t_w}{t}\exp\left(\Delta E_j \frac{T_{build}-T_{erasure}}{T_{build}T_{erasure}}\right)\right]$. The total erasure curve used to fit the data in Figure 7 and Figure 9 is the sum of such contributions with $\Delta E_j$ running from 0 to $\Delta E_{max}$, the upper bound being the fitting parameter.

Note that since the flat part of $N(m)$ in temperature independent, in the corresponding temperature range non isothermal TDE would be unable to reveal the *T* dependence of the dynamics.

**References**

[1] Pollak M, Ortuno M and Frydman A, *The Electron Glass* (Cambridge University Press, 2012)
Ovadyahu Z, *Comptes Rendus Physique* **14**, 637 (2013)
[2] Vaknin A, Ovadyahu Z and Pollak M, *Phys. Rev. Lett.* **84**, 3402 (2000)
[3] Havdala T, Eisenbach A and Frydman A, *EPL* **98**, 67006 (2012)
Adkins C J, Benjamin J D, Thomas J M D, Gardner J W and McGeown, *J Phys.* C: *Solid State Phys.*, **17**, 4633 (1984)
[4] Delahaye J, Grenet T, Marrache-Kikuchi C, Drillien A A and Berge L, *EPL* **106**, 67006 (2014)
[5] Grenet T and Delahaye J, *Eur. Phys. J.* B **76**, 229 (2010)
[6] Grenet T and Delahaye J, *Phys. Rev.* B **85**, 235114 (2012)
[7] Grenet T, Delahaye J, Sabra M and Gay F, *Eur. Phys. J.* B **56**, 183 (2007)
[8] Amir A, Oreg Y and Imry Y, *Phys. Rev. Lett.* **103**, 126403 (2009)
[9] Müller M and Pankov S, *Phys. Rev.* B **75**, 144201 (2007)
[10] Ovadyahu Z, *Phys. Rev.* B **78**, 195120 (2008)
[11] Ovadyahu Z, *Phys. Rev. Lett.* **99**, 226603 (2007)
[12] Vaknin A, Ovadyahu Z and Pollak M, *Phys. Rev.* B **65**, 134208 (2002)
[13] Vaknin A, Ovadyahu Z and Pollak M, *Phys. Rev. Lett.* **81**, 669 (1998)
[14] Ovadyahu Z, *Phys. Rev.* B **73**, 214208 (2006)